\documentclass[conference]{IEEEtran}
\IEEEoverridecommandlockouts
\usepackage{cite}
\usepackage{amsmath,amssymb,amsfonts}
\usepackage{algorithmic}
\usepackage{graphicx}
\usepackage{textcomp}
\usepackage{xcolor}
\usepackage{array, caption, tabularx,  ragged2e,  booktabs}
\usepackage{enumitem}
\usepackage{arabtex}
\usepackage{tabularray}
\usepackage{comment}
\usepackage{amsmath}
\usepackage[]{url}

\def\BibTeX{{\rm B\kern-.05em{\sc i\kern-.025em b}\kern-.08em
    T\kern-.1667em\lower.7ex\hbox{E}\kern-.125emX}}
\begin{document}

\title{The Passwordless Authentication with Passkey Technology from an Implementation Perspective\\
} 

\begin{comment}
\author{
\IEEEauthorblockN{Lien Tran}
\IEEEauthorblockA{ \textit{School of Computer Science \& Tech.} \\
\textit{Algoma University}\\
Sault Ste. Marie, ON, Canada \\
ltran@algomau.ca}\\

\IEEEauthorblockN{Boyuan Zhang}
\IEEEauthorblockA{ \textit{School of Computer Science \& Tech.} \\
\textit{Algoma University}\\
Sault Ste. Marie, ON, Canada \\
bzhang@algomau.ca}\\

\and

%\vspace{4.5mm}
\IEEEauthorblockN{Ratchanon Pawanja}
\IEEEauthorblockA{ \textit{School of Computer Science \& Tech.} \\
\textit{Algoma University}\\
Sault Ste. Marie, ON, Canada \\
rpawanja@algomau.ca}\\

%\vspace{4.5mm}
\IEEEauthorblockN{Rashid Hussain Khokhar}
\IEEEauthorblockA{ \textit{School of Computer Science \& Tech.} \\
\textit{Algoma University}\\
Sault Ste. Marie, ON, Canada \\
rashid.khokhar@algomau.ca}

}
\end{comment}

\author{
Lien Tran, Boyuan Zhang, Ratchanon Pawanja, and Rashid Hussain Khokhar \\
School of Computer Science \& Technology, Algoma University, Sault Ste. Marie, ON, Canada \\Emails: 
\{ltran, bzhang, rpawanja, rashid.khokhar\}@algomau.ca
}

\maketitle

\begin{center}
\textcolor{blue}{\textbf{Preprint.} This version has been accepted at 2nd IEEE/ACIS International Conference on Software Engineering, Artificial Intelligence, Networking and Parallel/Distributed Computing (SNPD2025-Summer IV).}
\end{center}

\begin{abstract}
With the rise of sophisticated authentication bypass techniques, passwords are no longer considered a reliable method for securing authentication systems. In recent years, new authentication technologies have shifted from traditional password-based logins to passwordless security. Among these, Time-Based One-Time Passwords (TOTP) remain one of the most widely used mechanisms, while Passkeys are emerging as a promising alternative with growing adoption. This paper highlights the key techniques used during the implementation of the authentication system with Passkey technology. It also suggests considerations for integrating components during system development to ensure that users can securely access their accounts with minimal complexity, while still meeting the requirements of a robust authentication system that balances security, usability, and performance. Additionally, by examining TOTP and Passkey mechanisms from an implementation perspective, this work not only addresses major security concerns such as password leaks, phishing attacks, and susceptibility to brute-force attacks, but also evaluates the feasibility and effectiveness of these mechanisms in real-world implementations. This paper demonstrates the superior security of Passkey technology and its potential for broader adoption in secure authentication systems.
\end{abstract}


\section{Introduction}
Since technology has continued to develop rapidly in recent years, severe cyberattacks have also increased dramatically and are becoming more challenging than ever to protect sensitive data and users from unauthorized access. Traditional password mechanisms are too simple and no longer sufficient to prevent sophisticated attacks.  To mitigate these risks, Time-Based One-Time Passwords (TOTP) and Passkey are superior alternatives in the authentication process to replace password-based systems. TOTP, a lightweight cryptographic algorithm and an advanced version of OTP, reduces computational load. However, it remains susceptible to brute-force attacks, which is a primary concern for this technology~\cite{Sofian2024enhancing}. In contrast, Passkey employs a public-private key mechanism, resulting in better phishing resistance. Furthermore, the primary advantage of Passkey technology lies in its integration of hardware tokens with built-in device capabilities~\cite{shankar2024passwordless}. Some of the key aspects of Passkey, such as multi-device synchronization, phishing resistance, and seamless user experience, will be discussed in the paper. 
%\subsection{Contributions}
With the deployment of passwordless authentication with passkey technology, our primary contributions include:
\begin{itemize}
\item{\textit{Proposed Framework:}} The proposed framework, constructed using JavaScript and Python, aims to provide developers with a foundational structure to build authentication systems that seamlessly integrate Passkey technology while retaining the familiar features of TOTP authentication, thereby ensuring a smooth transition and enhancing user adoption.

\item{\textit{Implementation Guidelines:}} This work outlines the key considerations for developers deploying authentication solutions in real-world environments. By evaluating the security characteristics of Passkey in practical deployments, this contribution provides a foundational framework to help organizations make informed decisions about adopting new authentication technologies. 
\end{itemize}

%\subsection{Organization}
%Add organization of the paper
The remainder of the paper is organized as follows: Section II reviews the relevant literature on TOTP and Passkey. Section III details the methodology. Section IV presents the experimental setup and offers recommendations for deployment. Section V provides a discussion, and Section VI concludes the paper.

\section{Literature Review}
To determine the optimal authentication mechanism for the deployment system, this paper evaluates the advantages and disadvantages, as well as the challenges in deploying TOTP and Passkey, based on existing research. Sofian et al.~\cite{Sofian2024enhancing} highlight that the use of AES and DES encryption standards in TOTP contributes to a secure mechanism for critical sectors. Due to its lightweight features, TOTP may also address some of the security limitations in IoT environments. However, Berrios et al.~\cite{Berrios23_Factorizing2FA} state that TOTP is not able to resist all adversaries. The primary vulnerability lies in plaintext secret keys. In the event of obtaining the key, an attacker can easily gain access without the original device. 

Concerning cybersecurity threats, the vulnerability of TOTP, as observed by Dixit et al.~\cite{dixit2024development}, also carries the risk of password exposure through phishing attacks. Although this weakness lies in the human element, TOTP itself does not inherently prevent users from revealing their static password and the one-time code when users are tricked into fraudulent sites. Additionally, the entropy limitations of TOTP are further corroborated in recent work by Nair and Song~\cite{nair2023mfkdf}. The 20-bit entropy of 6-digit TOTP codes, despite its popularity as an authentication factor, is not sufficient to resist brute-force attacks. Furthermore, Ding and Wang~\cite{10919149} also point out that in the 2FA framework, relying on the login terminal could expose a potential risk to users. They also suggested that it is more secure for users to exclusively trust their own devices.  

George~\cite{george2024passkeys} extensively discusses the vulnerabilities of passwords and positions Passkey as the optimal replacement. With WebAuthn and Fast Identity Online (FIDO) protocols, Passkey has limited the password transmitting process, leveraging public-key cryptography. Regarding the security aspect, Passkey has demonstrated robust cryptographic strength to resist phishing, man-in-the-middle, and replay attacks. Ou et al.~\cite{Ou2024} also state that Passkey offers users flexible control and convenience when logging into multiple devices, making it an advanced solution.

Nevertheless, the use of a passkey as a replacement for passwords still faces some challenges in practical scenarios. Matzen et al.~\cite{matzen2025challenges} have provided several possible hesitations among users who, due to a tendency to share their accounts or delegate access, would find the passkey inconvenient, given that the private key is stored on a specific device. Regarding account recovery, the authors also note the limitation of the recovery mechanism, which has not yet been standardized. Although both TOTP and Passkey have outstanding benefits and drawbacks, few studies provide practical deployment frameworks and deployment considerations to utilize the advantages of both methods.

\section{Methodology}
We developed a decoupled front- and back-end authentication system that supports multiple login methods, including user registration, email-based login with verification codes, Google OAuth authentication, and Passkey login to enhance security. The system integrates modern technologies, frameworks, and security protocols to create authentication. The system is managed, deployed, and run using Docker\footnote{Docker is an open-source platform for packaging applications and their dependencies into lightweight, portable containers.} to provide a secure and high performance authentication solution.\\

To achieve a robust and fortified security system, A comprehensive analysis of the TOTP and Passkey methods has been conducted to explore a hybrid implementation approach. Our implementation framework prioritizes Passkey as the primary authentication method while maintaining TOTP as a strategic backup. Fig. 1 illustrates the workflow and setup for Passkey and TOTP in our deployment.
\begin{figure}[h!]
  \centering
 \includegraphics[width=0.45\textwidth]{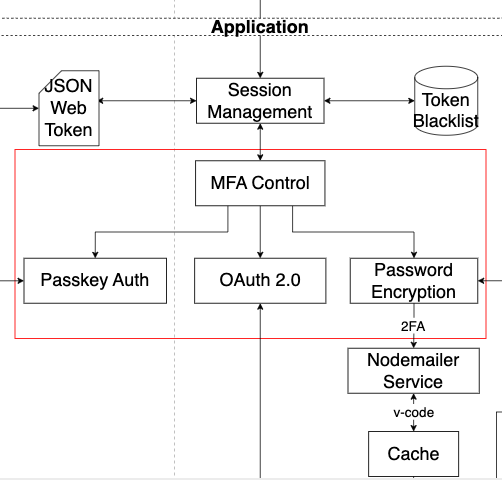}
  \caption{Passkey and TOTP Authentication Deployment}
  \label{fig:my_image1}
\end{figure}

\subsection{Passkey Security Characteristics}
\begin{itemize}
    \item \textit{Brute Force Protection:} Passkey uses a unique challenge-response mechanism without the presence of a password to effectively prevent brute-force attacks.  
    
    \item \textit{Phishing Prevention:} Since the private key is stored on the user's device and authentication is bound to specific domains, phishing attacks are effectively prevented even if users are tricked into fraudulent sites.
    
    \item \textit{Replay Attack Prevention:} A new random challenge is generated for each authentication attempt, and when combined with timestamped signatures, it helps neutralize replay attacks.
\end{itemize}

\subsection{TOTP Implementation}
During the registration phase, Two-Factor Authentication (2FA) via TOTP is implemented to enhance security facilitated by the speakeasy \footnote{https://github.com/speakeasyjs/speakeasy} library, which generates a unique secret key for the user using the $speakeasy.generateSecret()$ method. This secret key, associated with the user's account, allows them to verify the registration process.

Even though TOTP was selected during the registration process, our preference for login purposes is Passkey because of several concerns regarding the inconsistent time synchronization between the user's device and the server of TOTP. This inconsistency can lead to login failures, and more critically, a sophisticated attacker might exploit these timing discrepancies to generate and save valid TOTP codes corresponding to future time intervals on a user's device and bypass authentication~\cite{BianchiValeriani2023TOTP}.

\subsection{Passkey Implementation}
Passkey is a new authentication standard based on the WebAuthn specification developed by the FIDO Alliance and W3C, offering a convenient authentication experience for users~\cite{hal-passkey2020}. Passkey is implemented in both the back-end and front-end components to ensure a seamless passwordless experience for users.

A combination of modern technologies, frameworks, and security protocols is used to create a resilient authentication system. The system is managed, deployed, and run using Docker, starting with basic features and progressing to more advanced techniques.

\subsection{Web App Security}
This paper employs two key technologies: Bcrypt\footnote{https://www.npmjs.com/package/bcryptjs} and JSON Web Tokens (JWT)\footnote{https://jwt.io/} to improve security and performance. Bcrypt is a prominent hashing function used for securing passwords. Created by Niels Provos and David Mazieres based on the Blowfish algorithm, Bcrypt incorporates a salt to generate a unique hash, converting plain-text passwords into a fixed-length, non-reversible format. Compared to SHA-256 hashing algorithms, Bcrypt is outperformed in the event of brute-force attacks~\cite{skanda2022secure}. Besides, with an advanced characteristic of work factor, this algorithm can determine the resources needed to hash a password according to the computational resources~\cite{kolhe2024designAJBR}. Fig. 2 represents the process of storing passwords using salted hashing with bcrypt. The cost factor (work factor) is configured as follows.
\begin{verbatim}
bcrypt.hash(password, 10)
\end{verbatim}
The number 10 represents the salt rounds, meaning bcrypt performs $2^{10}$ iterations to generate the hash. Each password gets a unique salt automatically.

\begin{verbatim}
bcrypt.compare(oldPassword, user.password)
\end{verbatim}

$bycrypt.compare()$ is the function to examine the old password with the new password by re-hashing the provided password using the salt from the stored hash. It returns $true$ if the hashes match.\\

\vspace{-0.5cm}
\begin{figure}[!ht]
  \centering
\includegraphics[width=0.5\textwidth]{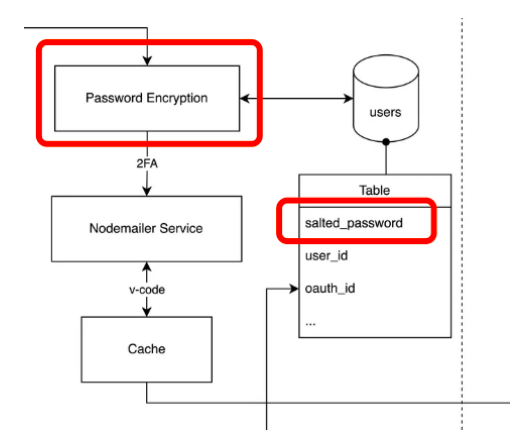}
  \caption{Password Encryption with Salting}
  \label{fig:my_image2}
\end{figure}

JSON Web Tokens (JWT) provide a foundational framework for authentication and access control in modern web applications. With JWT-based authentication, relevant user access information is embedded into a token and subsequently signed using the JWT secret. Although this approach offers scalability, a major drawback of JWT is the difficulty of revoking tokens at will~\cite{JWT2018}. To address this, a blacklisting mechanism was implemented  to validate each token included in incoming requests. A token blacklist allows the system to revoke JWTs before their expiration, helping prevent potential compromise. \\
\begin{verbatim}
token: {
    type: String,
    required: true,
    unique: true,
  },
  createdAt: {
    type: Date,
    default: Date.now,
    expires: "1h", 
  }
\end{verbatim}
This setup provides a simple blacklisting mechanism using a Mongoose schema, which corresponds to a MongoDB\footnote{MongoDB is a NoSQL database that stores data JSON-like documents} collection named ``TokenBlacklist" Schema.\\
The field named \texttt{token} of type `String' is intended to hold the actual JWT or a unique identifier of a revoked token.\\
The field named \texttt{createdAt} of type `Date' represents a JavaScript `Date' object.\\
Mongoose will automatically set its value to the current date and time with \texttt{default:} $Date.now()$. MongoDB will automatically delete documents from the ``TokenBlacklist" collection one hour after creation, as defined by \texttt{expires:} ``1h". \\

\subsection{Implementation Details}
\label{subsec:implementation}

\underline{\textbf{Basic Components}}:
\begin{itemize}
    \item \textit{Password Encryption}: bcrypt (for password hashing and verification)
    
    \item \textit{JWT Implementation}:
    \begin{itemize}
        \item Uses \texttt{jsonwebtoken} library for token generation
        \item Token structure: \texttt{header.payload.signature}
        \begin{itemize}
            \item Header: Token type and algorithm
            \item Payload: Encoded data
            \item Signature: Generated via \texttt{sign()} method using header, payload, and secret key
        \end{itemize}
    \end{itemize}
    
    \item \textit{Token Blacklist}:
    \begin{itemize}
        \item MongoDB query: \texttt{TokenBlacklist.findOne(\{ token \})}
        \item Returns ``Token has been revoked" if found
    \end{itemize}
\end{itemize}

\underline{\textbf{Advanced Components}}:
\begin{itemize}
    \item \textit{Passkey Authentication}:
    \begin{itemize}
        \item Passwordless auth using WebAuthn standard
        \item Native browser APIs + \texttt{@simplewebauthn/ server} (Node.js backend)
        \item Key files: \texttt{passkeyController.ts}
    \end{itemize}
    
    \item \textit{Storage and Process Management}:
    \begin{itemize}
        \item \texttt{PasskeyCredential}: Stores credentials (ID, public key, counter, device name)
        \item \texttt{PasskeySession}: Tracks registration/authentication status
    \end{itemize}
    
    \item \textit{API Endpoints}:
    \begin{itemize}
        \item Registration: \texttt{/auth/passkey/register-
        options}, \texttt{/auth/passkey/register-verify}
        \item Authentication: \texttt{/auth/passkey/auth-options}, \texttt{/auth/passkey/auth-verify}
        \item Management: \texttt{/auth/passkey/list}, \texttt{/auth/passkey/:id}
    \end{itemize}
    
    \item \textit{2FA}: Nodemailer (verification code delivery)
    
    \item \textit{OAuth 2.0}: Passport.js (Google OAuth integration)
\end{itemize}

\section{Experimental Setup}
The experiment will demonstrate various authentication approaches, including password-based authentication, passkeys, two-factor authentication (2FA), and OAuth2 third-party login, with full-stack implementation. The source code, along with deployment scripts, is available on GitHub\footnote{https://github.com/Alex-xd/login-system}.

\subsection{Architectural Design}
To support multiple secure login methods, including user registration, email-based verification, Google OAuth authentication, and Passkey login, the system employs a modular architecture. This design, facilitated by Docker, also ensures consistent deployment environments in different phases of the implementation. In addition, the isolation capabilities of this design help reduce security risks in the event of a compromise.

\subsection{Technologies}

\begin{description}[labelindent=0pt, leftmargin=!]
  \item \textit{Framework:} Node.js, Express
  \item \textit{Language:} TypeScript
  \item \textit{Database:} MongoDB (via Mongoose ODM)
  \item \textit{Cache:} Redis (for storing short-term verification codes)
  \item \textit{Authentication Framework:} Passport.js (for Google OAuth 2.0 authentication)
  \item \textit{WebAuthn:} \texttt{@simplewebauthn/server} (for Passkey authentication)
  \item \textit{Email Service:} Nodemailer (for sending verification codes)
  \item \textit{Two-Factor Authentication (2FA):} speakeasy (for generating and verifying TOTP tokens)
  \item \textit{Password Encryption:} bcrypt (for password hashing and verification)
  \item \textit{Security Middleware:} helmet (for HTTP header security)
  \item \textit{Session Management:} express-session
  \item \textit{Rate Limiting:} express-rate-limit (to prevent brute force attacks)
  \item \textit{Token Generation:} jsonwebtoken (JWT implementation)
  \item \textit{Environment Variables:} dotenv
\end{description}

\subsection{Logic Implementation}
\begin{enumerate}
    \item \textit{User Registration}: Once the user enters their email on the registration page and requests a verification code, the system stores the generated code in Redis with a 30-second expiration and saves the email in a temporary user table. Nodemailer is then used to send the verification email using predefined configurations. If the entered email and code match the stored values, the user is allowed to set a password. Upon successful verification, the system updates the temporary table (setting `otpVerified' to $true$) and deletes the verification code from Redis. The entered password is then hashed and stored in the temporary table. Finally, the user’s information is saved in the MongoDB database, and the corresponding entry in the temporary table is deleted. The system returns a message indicating whether the registration was successful or failed.

\item \textit{Email and Verification Code Authentication Flow}: On the login page, the user enters their email and password. If the credentials are valid, the system generates a verification code and sends it to the user’s email. Upon successful verification of the code, the system generates a JWT token, stores it in localStorage, and redirects the user to the profile page. Once the authorization code for an access token is exchanged and the user’s Google profile information is retrieved, the system checks whether the user already exists in MongoDB. If the user exists, their information is returned; otherwise, a new user is created and stored in the database. Finally, the system generates a JWT and returns it to the frontend along with a response indicating whether the operation was successful or failed.

\item \textit{Google Login}: The user clicks the ``Google Login” button, which redirects them to Google’s OAuth 2.0 authorization page. There, Google prompts the user to grant permission to access their account information. Once an authorization code is obtained, Google redirects the user back to the backend, where the system uses Passport.js’ GoogleStrategy to handle the OAuth 2.0 callback.

\item \textit{Passkey Registration}: Upon the user requesting Registration Options, the backend receives the request and generates WebAuthn registration options, including a challenge value and user information. Next, the system creates a session record, saves the challenge, and sets an expiration time. The backend then returns the options and session ID to the frontend, which will later call $navigator.credentials.create()$ with the provided options and prompt the user for biometric authentication (e.g., Face ID) or a security key. After authentication, the browser generates credentials and sends them, along with the device name, to the backend. The system retrieves the session record and verifies the credentials using WebAuthn libraries. If the credentials are valid, the backend stores them in the database and marks the session as completed.

\item \textit{Passkey Login Process}: When the user clicks ``Login with Passkey,” the frontend sends a request for authentication options. The backend receives this request and creates WebAuthn authentication options, including a challenge value, and temporarily stores the session record and challenge. Subsequently, the frontend calls $navigator.credentials.get()$ with the provided options, prompting the browser for biometric authentication. Fig. 3 illustrates the Passkey login flow. After successful authentication, the browser submits the authentication response to the backend. The backend then verifies the response using WebAuthn libraries. If the response is valid, the backend updates the credential counter to prevent replay attacks, generates a JWT token, and sends it to the frontend along with a successful response. Finally, the frontend saves the JWT token and redirects the user to the homepage.
\end{enumerate}
\vspace{-0.5cm}
\begin{figure}[h!]
  \centering
  \includegraphics[width=0.5\textwidth]{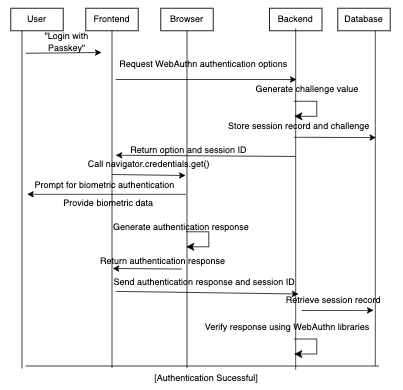}
  \caption{Passkey Login Authentication Sequence Map}
  \label{fig:my_image}
\end{figure}

%\end{itemize}

\subsection{Results}

%\begin{description}[labelindent=0pt, leftmargin=!]
\begin{itemize}
  \item \textit{Password-Based Authentication:} The secured password hashing using bcrypt was successfully implemented to avoid common attacks like brute force or dictionary attacks.
  
  \item \textit{Passkey Authentication:} Integrating Passkey authentication using WebAuthn (\texttt{@simplewebauthn/server}) eliminated the need for passwords and significantly reduced the risk of phishing attacks and credential theft.
  
  \item \textit{Two-Factor Authentication (2FA):} Implemented email-based 2FA using Nodemailer. A unique verification code was sent whenever a user logged in and must be entered to complete the process. If the password was compromised, access would be denied without the verification code.
  
  \item \textit{OAuth2 Integration:} OAuth2 authentication was integrated through Passport.js to enable Google login. The implementation also included token handling and permission scopes to ensure user data protection.
  
  \item \textit{JWT Authentication:} Implemented secure token-based authentication using jsonwebtoken (JWT). A predefined session duration in which users were logged in would automatically end when the tokens expired. In the event of a security breach, a token blacklist would revoke the token.
  
  \item \textit{Rate Limiting:} Rate limiting using express-rate-limit was added to prevent brute-force attacks and enhance overall system security. With rate limiting, the risk of automated attacks is significantly minimized, leading to a more secure and reliable authentication system.
  
  \item \textit{HTTP Header Security:} Security middleware (helmet) was incorporated to strengthen HTTP headers, reducing vulnerabilities to common web attacks.
  
  \item \textit{Caching:} Redis was used to cache temporary data and significantly improved authentication performance by reducing database load and speeding up data retrieval.
  
  \item \textit{Session Management:} Session management was handled via express-session, ensuring secure and efficient user session handling.

  \item \textit{Database:} MongoDB was used as the database, with Mongoose ODM facilitating structured interaction.
  
  \item \textit{Environment Variables:} Configuration parameters were managed securely using dotenv.
\end{itemize}  
%\end{description}

\subsection{Implementation Considerations}
Several factors should be taken into account when implementing authentication using passkeys and TOTP. Firstly, platform support is vital for this work; developers must check browser compatibility with the WebAuthn API and consider fallback authentication methods for unsupported browsers. For compatibility, legacy devices may not support passkeys, and certain environments might restrict their usage. As mentioned above, the recovery mechanism has a significant impact on user experience; for that reason, ensuring a robust recovery and providing a clear on-boarding process would help users to adapt to the new system. The FIDO Alliance guidelines\footnote{https://www.passkeycentral.org/design-guidelines/} and Passkey Roll-Out Guides\footnote{https://www.passkeycentral.org/passkey-roll-out-guides} provide a comprehensive foundation that developers can reference to follow security best practices tailored to business needs, such as proper key storage on the server side, rate limiting, and secure session management. Finally, it is important to handle edge cases to ensure a smooth and secure user experience.

\subsection{Lessons Learned}
One of the most significant lessons learned during the development of this secure authentication system is the complexity involved in integrating various components and ensuring they work seamlessly together. The integration of multiple security methods requires careful consideration of their interdependencies, which are reflected in the system's modular architecture, as illustrated by the distinct directories for controllers, models, routes, and middlewares. Additionally, conflicts can occur during the process of configuration fine-tuning, and inappropriate settings can lead to overly restricted or insufficient security, potentially affecting the effectiveness of the authentication system.

\subsection{Recommendations}
Based on our deployment experience and comparative analysis, this work proposes the following strategies to ensure successful implementation. Given the limitations of TOTP compared to passkeys, adopting passkeys as the primary authentication method will significantly enhance system security. TOTP can serve as a suitable backup or supplementary mechanism. This layered approach supports a smooth user transition and reinforces a clear and transparent communication policy regarding authentication practices. For high-security scenarios, such as financial transactions, enforcing passkey-only authentication is recommended. For general use cases, an optional TOTP fallback mechanism, accompanied by appropriate risk warnings, should be sufficient. In special cases, it may be reasonable to allow flexible configurations, such as temporary TOTP support for legacy systems. To ensure a smooth transition, passkey authentication should be rolled out gradually. Maintaining TOTP support during the migration phase is essential, along with monitoring user acceptance and adjusting policies based on feedback. 

\section{Discussion}
TOTP is a widely used, lightweight two-factor authentication mechanism. However, it relies on a shared secret between the client and the server, which introduces inherent security vulnerabilities. If the secret is compromised, the authentication process can be bypassed, and the need for manual code entry may introduce additional usability friction. In contrast, Passkey authentication, implemented via the WebAuthn standard, offers a better experience and a more secure method of authentication. It completely eliminates passwords by using asymmetric cryptography. Authentication occurs through the signing of a cryptographic challenge, typically using biometric input~\cite{george2024passkeys}. Recent industry trends reflect a shift toward the exclusive use of passkeys. For example, Porkbun adapts passkey-only authentication at the time of account creation~\cite{daffallaframework}. If an attacker gains access to a user’s email or password, authentication is still impossible without biometric verification on the user’s device. From a strict security perspective, passkeys provide stronger guarantees than TOTP. Under ideal conditions, a passkey-only system is sufficient to secure user accounts. Considering the redundancy between passkeys and TOTP, it is evident that passkeys offer a more robust and efficient solution for secure authentication. While TOTP may still play a role due to limited device compatibility with passkeys, its necessity diminishes as passkey adoption increases.

\section{Conclusion}
This work highlights the advantages of Passkey authentication over traditional password-based systems and TOTP-based two-factor authentication. Passkey’s reliance on asymmetric cryptography and biometric verification provides more security while minimizing user friction. However, while TOTP is still relevant in some contexts, it is increasingly becoming redundant as passkey adoption grows. This paper concludes that passkeys have strong potential to become the primary authentication method in future systems, offering a more secure and efficient approach to authentication. TOTP may still serve as a supplementary backup in scenarios where passkey implementation is not yet feasible.

\bibliographystyle{IEEEtran}
%\bibliography{conference_SS4-5}
%\input{conference_SS4-5.bbl}

%\vspace{12pt}

\end{document}